\documentclass[aip,reprint]{revtex4-2}
\usepackage{graphicx}
\usepackage{dcolumn}
\usepackage{bm}
\usepackage[utf8]{inputenc}
\usepackage[T1]{fontenc}
\usepackage{mathptmx}
\usepackage{mathrsfs}
\usepackage{gensymb}
\usepackage{amsfonts}
\usepackage{amsthm}
\usepackage{amssymb}
\usepackage{amsmath}
\usepackage{natbib}
\usepackage{color}
\usepackage{hyperref}
\usepackage{bm}
\usepackage[caption=false]{subfig}
\usepackage{verbatim}
\usepackage{ulem}
\usepackage{soul}

\DeclareUnicodeCharacter{0308}{\"}

\usepackage{xcolor}
\setstcolor{red}

\begin{document}

\title{Crystallographic Orientation-Dependent Magnetotransport in the Layered Antiferromagnet - CrSBr}
\author{Naresh Shyaga}
\affiliation{Centre for Nanoscience and Engineering, Indian Institute of Science, Bengaluru 560012, Karnataka, India} 
\author{Pankaj Bhardwaj}
\affiliation{Centre for Nanoscience and Engineering, Indian Institute of Science, Bengaluru 560012, Karnataka, India} 
\author{Rajib Sarkar}
\affiliation{Centre for Nanoscience and Engineering, Indian Institute of Science, Bengaluru 560012, Karnataka, India} 
\author{Jagadish Rajendran}
\affiliation{Centre for Nanoscience and Engineering, Indian Institute of Science, Bengaluru 560012, Karnataka, India}
\author{Abhiram Soori}
\affiliation{School of Physics, University of Hyderabad, Prof. C. R. Rao Road, Gachibowli, Hyderabad -- 500046, India} 
\author{Dhavala Suri}
\email{dsuri@iisc.ac.in}
\affiliation{Centre for Nanoscience and Engineering, Indian Institute of Science, Bengaluru 560012, Karnataka, India} 

\begin{abstract}
Among two-dimensional magnetic materials, CrSBr has attracted considerable attention owing to its coexistence of ferromagnetic and antiferromagnetic ordering, which depends sensitively on crystallographic orientation. An additional distinguishing feature of CrSBr is its highly anisotropic Fermi surface in momentum space. In this work, we present a comprehensive investigation of magnetoresistance by systematically orienting the bias current and the applied magnetic field along all three crystallographic axes. We demonstrate that the magnetoresistance serves as a direct probe of electronic anisotropy, exhibiting pronounced variations when the current is applied along different crystallographic directions under a magnetic field perpendicular to the sample plane. For in-plane magnetic fields, we observe conventional anisotropic magnetoresistance accompanied by hysteresis, indicative of ferromagnetic behavior.  Overall, our study provides a complete picture of electronic transport in CrSBr as a function of bias current and magnetic field orientation with respect to crystallographic directions, thereby opening pathways for future experiments requiring high sensitivity of electrical resistance to magnetic field gradients.
\end{abstract}
	
\maketitle


Chromium sulfide bromide (CrSBr) has recently attracted significant attention as a member of the emerging class of two-dimensional (2D) van der Waals (vdW) magnetic semiconductors \cite{lee_2021,liu_2023,wilson_2021,Rudenko_2023}. In contrast to conventional 2D materials that are either nonmagnetic or metallic, CrSBr uniquely combines intrinsic magnetism with semiconducting behavior, while maintaining excellent environmental stability \cite{ye_2022,ziebel_2024,Long_2023}. This combination positions CrSBr as a promising platform for exploring low-dimensional magnetism as well as for potential applications in spintronics, magneto-optoelectronics, and vdW heterostructure devices \cite{wang_2022,Liu_2025,Badola_2026,Khan_2024}.

CrSBr crystallizes in an orthorhombic layered structure, where each monolayer consists of a buckled Cr-S plane sandwiched between bromine atoms \cite{Beck_1990,ziebel_2024,Li_2025}. Strong covalent and ionic bonding within the layer coexists with weak interlayer vdW interactions, enabling mechanical exfoliation down to the monolayer limit \cite{Guo_2018,Telford_2022}. Magnetically, CrSBr exhibits pronounced anisotropy and A-type antiferromagnetic order, characterized by ferromagnetic coupling within individual layers and antiferromagnetic coupling between adjacent layers \cite{Goser_1990,lee_2021,Telford_2022}. The antiferromagnetic transition occurs at a N\'eel temperature $T_N \approx 130$--$140$~K, depending on sample quality and measurement conditions \cite{lopez_2022,Long_2023,Pei_2024}. Below $T_N$, long-range magnetic order is stabilized, while above this temperature the system transitions into a paramagnetic state with strong spin fluctuations \cite{rizzo_2022,scheie_2022,liu_2022}. A growing body of literature has investigated the electronic and optical properties of CrSBr, revealing a semiconducting band structure with a band gap in the visible to near-infrared range \cite{Guo_2018,Moros_2023,Mosina_2024,Biktagirov_2025}.  Optical spectroscopy studies have demonstrated pronounced excitonic features and strong linear dichroism, reflecting the underlying crystallographic and magnetic anisotropy \cite{lin_2024,mondal_2025,wilson_2021,Smiertka_2026}. Importantly, the coupling between magnetic order and electronic structure has been shown to give rise to magnetically tunable optical responses, making CrSBr an appealing candidate for magneto-optical and spin-dependent optoelectronic applications \cite{gong_2019,rahman_2021,Yang_2025,Chen_2025}.

Recent experimental and theoretical efforts have further focused on understanding charge transport and magneto-transport phenomena in CrSBr\cite{Hong_2025,Liu2_2025,Telford_2020,Boix_2022}. Strong anisotropy in electrical conductivity and magnetoresistance has been reported, arising from its highly anisotropic Fermi surface and spin-dependent scattering mechanisms \cite{Wang_2023,Henr_2025}. Additionally, the sensitivity of transport properties to magnetic field orientation has highlighted the intricate interplay between crystal symmetry, spin texture, and electronic states \cite{Wu_2022,Klein_2022}. Since the magnetic ordering depends critically on the axes of reference i.e., the ordering is anti-ferromagnetic along the z-axis, ferromagnetic along the x-axis and y-axis with preferred orientation along x-axis, charge transport exhibits strong magneto-transport anisotropy \cite{Wang_2023,Yang_2021,LopezPaz_2022}. Charge transport along different crystallographic axes exhibits a marked dependence on the relative orientation of the applied magnetic field. In particular, the transport response differs significantly when the magnetic field is aligned with the current direction compared to configurations in which it is not. In this work, we demonstrate that systematic electrical measurements under controlled electric and magnetic field orientations that provide direct insight into the anisotropy of the Fermi surface, thereby enabling a first-order electrical signatures of the underlying band structure. We show that straightforward magnetoresistance measurements can be employed to extract such information, offering a qualitative understanding of the Fermi surface morphology without resorting to more complex spectroscopic techniques.

\begin{figure*}[tbh!]
\includegraphics[width=\textwidth]{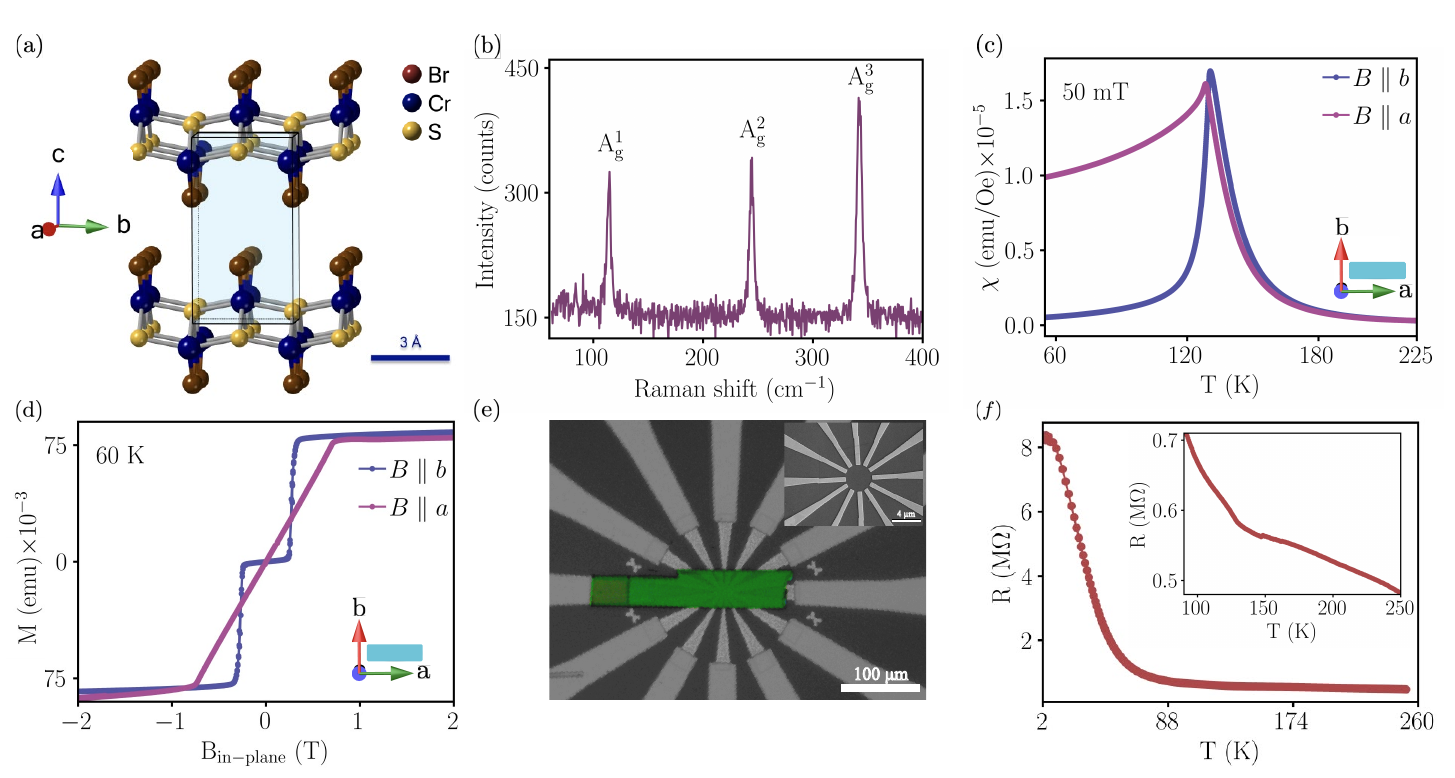}
\caption{(a) Schematics of the crystal structure of CrSBr. (b) Raman spectrum of a few-layer CrSBr flake. (c) Temperature-dependent susceptibility ($\chi$) at 50~mT applied along the a  and b directions of the crystal. (d) Field-dependent magnetization along a and b crystal axes at 60 K. (e) Optical micrograph of the transferred CrSBr flake onto the circular electrodes. False colour is applied to the flake for more visibility. Inset is the scanning electron microscopic image of the circular electrodes. (f) Resistance (R) as a function of temperature (T), and the inset is the zoomed-in region from 250 K to 100 K.}
\label{Fig1}
\end{figure*}

 Bulk CrSBr crystals were purchased from HQ Graphene and mechanically exfoliated onto SiO$_2$ (280~nm)/Si substrates using the Scotch tape method [the crystal structure is shown in Fig. 1 (a)]. Raman measurements were performed at room temperature on the 
exfoliated flakes using a 532~nm laser excitation to probe the vibrational modes and assess crystal quality. The Raman spectrum, shown in Fig.~1(b), exhibits three prominent peaks at 116.5~cm$^{-1}$, 247.6~cm$^{-1}$, and 345.9~cm$^{-1}$, corresponding to the  A$^1_{g}$, A$^2_{g}$, and A$^3_{g}$  vibrational modes, respectively\cite{sahu_2025}. These modes arise from symmetric optical phonons at the $\Gamma$ point, involving atomic vibrations perpendicular to the layer plane\cite{pawbake_2023}. The observed peak positions are in good agreement with previously reported values for CrSBr, and the sharpness of the peaks reflects the high crystalline quality of the flakes.

\noindent The magnetic properties of bulk CrSBr single crystals were investigated using vibrating sample magnetometry (VSM). Figure~\ref{Fig1}(c) shows the temperature dependence of the magnetic susceptibility ($\chi$) measured under an applied magnetic field of 0.05~T in the field-cooled configuration, with the field oriented along the crystallographic $a$- and $b$-axes. For both field orientations, $\chi$ increases upon cooling from room temperature and reaches a maximum, followed by a pronounced decrease below a characteristic temperature, indicative of an antiferromagnetic (AFM) phase transition. The temperature at which the cusp appears is identified as the N\'eel temperature, $T_N \approx 132$~K, in agreement with previous reports~\cite{Long_2023}. Below $T_N$, CrSBr exhibits antiferromagnetic interlayer coupling while maintaining ferromagnetic order within each layer. Above $T_N$, the system enters a paramagnetic regime, and long-range ferromagnetic order within the layers is absent. A marked anisotropy is observed in the susceptibility below $T_N$, with a steeper reduction for the field applied along the $b$-axis compared to the $a$-axis, highlighting the intrinsic magnetic anisotropy of CrSBr. This anisotropy is further corroborated by the field-dependent magnetization measurements performed at 60~K, as shown in Fig.~\ref{Fig1}(d). The magnetization ($M$) as a function of the applied magnetic field along both the $a$- and $b$-axes reveals a field-induced transition from the AFM to a ferromagnetic (FM) state, evidenced by the saturation of $M$ beyond a critical field. The saturation field ($B_{\mathrm{sat}}$) is strongly axis dependent, with a lower value of approximately 0.5~T along the $b$-axis compared to about 1~T along the $a$-axis, identifying the $b$-axis as the magnetic easy axis. For the field applied along the $b$-axis, the AFM--FM transition is sharp and exhibits a distinct double-step feature in the magnetization, which can be attributed to a spin-flip transition\cite{Telford_2020}. In contrast, when the field is applied along the $a$-axis, the magnetization increases more gradually with increasing field before reaching saturation, consistent with a spin-canting mechanism. We observe a decrease in the  $B_{sat}$ field with increasing temperature upto T$\mathrm{_N}$ for both crystal axes(refer to SI for details). Above T$\mathrm{_N}$, the M-B curves exhibit linear behaviour, corresponding to a paramagnetic (PM) nature of CrSBr. In addition, the temperature dependent susceptibility under various magnetic fields for both $a-$ and $b-$ axes demonstrates a transition from the (PM)-AFM state to the PM-FM state for the fields above 1~T and 0.5~T, respectively (see SI). These results confirm the strong magnetic anisotropy and layered antiferromagnetic nature of bulk CrSBr.

\noindent Circular electrodes were patterned by electron-beam lithography, as shown in the inset of Fig.~\ref{Fig1}(e). The electrodes were fabricated by depositing Ti (10~nm)/Au (50~nm) on a SiO$_2$ (280~nm)/Si substrate, followed by a lift-off process. CrSBr flakes were mechanically exfoliated onto a polydimethylsiloxane (PDMS) stamp, and a selected flake was subsequently transferred onto the circular electrodes using a micromanipulator-based alignment stage, as illustrated in Fig.~\ref{Fig1}(e). The exfoliated flakes (SI for details) are predominantly rectangular in shape, which arises from anisotropic crystal growth, wherein one crystallographic axis grows faster than the others. In addition, CrSBr crystallizes in an orthorhombic structure, characterized by unequal lattice parameters with interaxial angles of $90^{\circ}$. We measured the temperature dependence of the resistance \(R(T)\) of a transferred CrSBr flake, as shown in Fig.~\ref{Fig1}(f). The resistance increases monotonically over the entire temperature range, consistent with the semiconducting character of CrSBr. A pronounced upturn in \(R(T)\) appears near \(T \approx 129~\text{K}\), which coincides with the onset of antiferromagnetic (AFM) ordering, as highlighted in the inset of Fig.~\ref{Fig1}(f). This magnetic transition is further evidenced by a distinct local minimum in the first derivative \(dR/dT\) at the same temperature (see Supplementary Information), which can be attributed to the crossover from the paramagnetic to the AFM phase.

\begin{figure}[tbh!]
\includegraphics[width=\columnwidth]{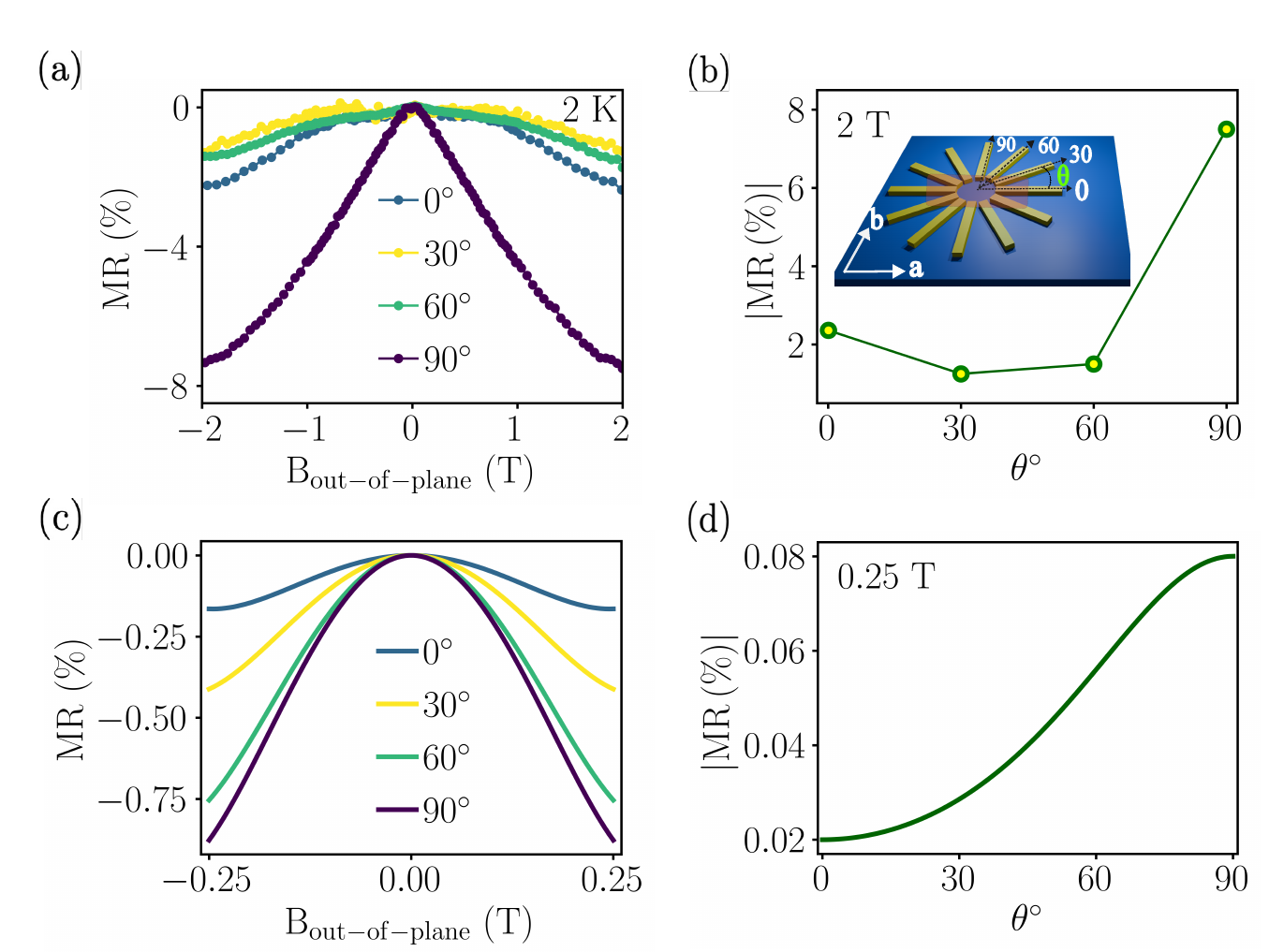}
\caption{(a) Angle-resolved magnetoresistance (MR) at 2 K under an out-of-plane magnetic field, where the current is applied along various directions with respect to the crystal axis-a (0$^\circ$); (thickness $\approx$  nm)  (b) Angle-dependent |MR| obtained for 2 T. Phenomenological model given by eqn (2) for MR (c) MR versus B$_{\text{out-of-plane}}$ (d) MR versus $\theta$.}
\label{Fig2}
\end{figure}

We study the magnetoresistance of the device when magnetic field is applied perpendicular to the sample plane; here the device electrodes are patterned in a circular manner to span various angles w.r.t. crystallographic axes.  Circular electrode patterns were fabricated by electron-beam lithography, as described for the inset of Fig.~\ref{Fig1}(e). Exfoliated CrSBr flakes were transferred onto the patterned substrate such that the crystallographic $a$-axis was aligned with one pair of electrodes, which we define as $0^{\circ}$. Additional electrode pairs were positioned at angular offsets of $30^{\circ}$, $60^{\circ}$, and $90^{\circ}$ with respect to the $a$-axis, with the $90^{\circ}$ configuration corresponding to the $b$-axis of CrSBr. The sample was mounted in an out-of-plane magnetic field configuration, and magnetoresistance was measured for different in-plane current directions by sweeping the applied magnetic field $B$. The magnetoresistance is defined as
\begin{equation}
MR(\%) = \frac{R(B) - R(0)}{R(0)} \times 100,
\end{equation}
where $R(B)$ is the resistance under an applied magnetic field and $R(0)$ is the zero-field resistance. As shown in Fig.~\ref{Fig2}(a), the magnetoresistance (MR) is negative for all current orientations, which can be attributed to the suppression of spin-disorder scattering upon application of a magnetic field. A pronounced enhancement of MR is observed for the $90^{\circ}$ configuration compared to the other angles. To quantify this anisotropy, we extracted the absolute value of MR at $B = 2$~T for all current directions, as plotted in Fig.~\ref{Fig2}(b). The MR values along $0^{\circ}$ (a-axis), $30^{\circ}$, $60^{\circ}$, and $90^{\circ}$ (b-axis) are 2.3\%, 1.3\%, 1.7\%, and 7.5\%, respectively. The MR is maximal along the $90^{\circ}$ direction and minimal at $30^{\circ}$. While the MR along $0^{\circ}$ exceeds that at $30^{\circ}$ and $60^{\circ}$, it remains significantly smaller than that along $90^{\circ}$. This systematic angular dependence of MR demonstrates the pronounced in-plane anisotropy of CrSBr.

To phenomenologically model the angular dependence of the magnetoresistance, we assume that the sample exhibits different longitudinal resistances along two orthogonal in-plane directions, denoted by $R_x$ and $R_y$. Such an anisotropy can arise from the underlying electronic bandstructure~\cite{soori2025}, but no microscopic mechanism is assumed here. Provided that the sample geometry is approximately identical for resistance measurements along the two directions, the resistance measured at an in-plane rotation angle $\theta$ (measured from the $x$-direction) in the presence of a perpendicular magnetic field $B$ can be written as
\[
R(B,\theta) = R_x(B)\cos^2\theta + R_y(B)\sin^2\theta,
\]
which represents the simplest phenomenological form consistent with two fold in-plane anisotropy~\cite{McGuire_1975}.

The magnetic-field-induced change in resistance along each direction is defined as
$\Delta R_{x/y}(B) = R_{x/y}(B) - R_{x/y}(0)$, leading to
\[
\Delta R(B,\theta) = \Delta R_x(B)\cos^2\theta + \Delta R_y(B)\sin^2\theta.
\]
The MR is therefore given by
\begin{equation}
MR(B,\theta) =
\frac{\Delta R_x(B)\cos^2\theta + \Delta R_y(B)\sin^2\theta}
{R_x \cos^2\theta + R_y \sin^2\theta}\times 100
\label{eq:model}
\end{equation}

From the experimental data, the low-field dependence of $\Delta R_{x/y}(B)$ is well captured by a polynomial expansion,
$\Delta R_{x/y}(B) = \gamma_{2,x/y} B^2 + \gamma_{4,x/y} B^4$,
which we use purely as a fitting form. Fitting the $\Delta R_{x/y}(B)$ versus $B$ data in the range
$B \in (-0.25~\mathrm{T},\,0.25~\mathrm{T})$ for $\theta=0^{\circ}, 90^{\circ}$ yields
$\gamma_{2,x} = -0.1923~\mathrm{M}\Omega/\mathrm{T}^2$,
$\gamma_{4,x} = 1.582~\mathrm{M}\Omega/\mathrm{T}^4$,
$\gamma_{2,y} = -1.167~\mathrm{M}\Omega/\mathrm{T}^2$, and
$\gamma_{4,y} = 6.305~\mathrm{M}\Omega/\mathrm{T}^4$.
The zero-field resistances extracted directly from the data are
$R_x = 3.459~\mathrm{M}\Omega$ and $R_y = 5.509~\mathrm{M}\Omega$.
Using these experimentally determined parameters in Eq.~\eqref{eq:model}, we obtain the calculated angular and field dependence of the MR shown in Fig.~\ref{Fig2}(c,d). The discrepancy in the nature of the curves obtained theoretically and the experimental data stem from the approximations in theory that is highly generic and not specific to CrSBr.

\begin{figure}[tbh!]
\includegraphics[width=\columnwidth]{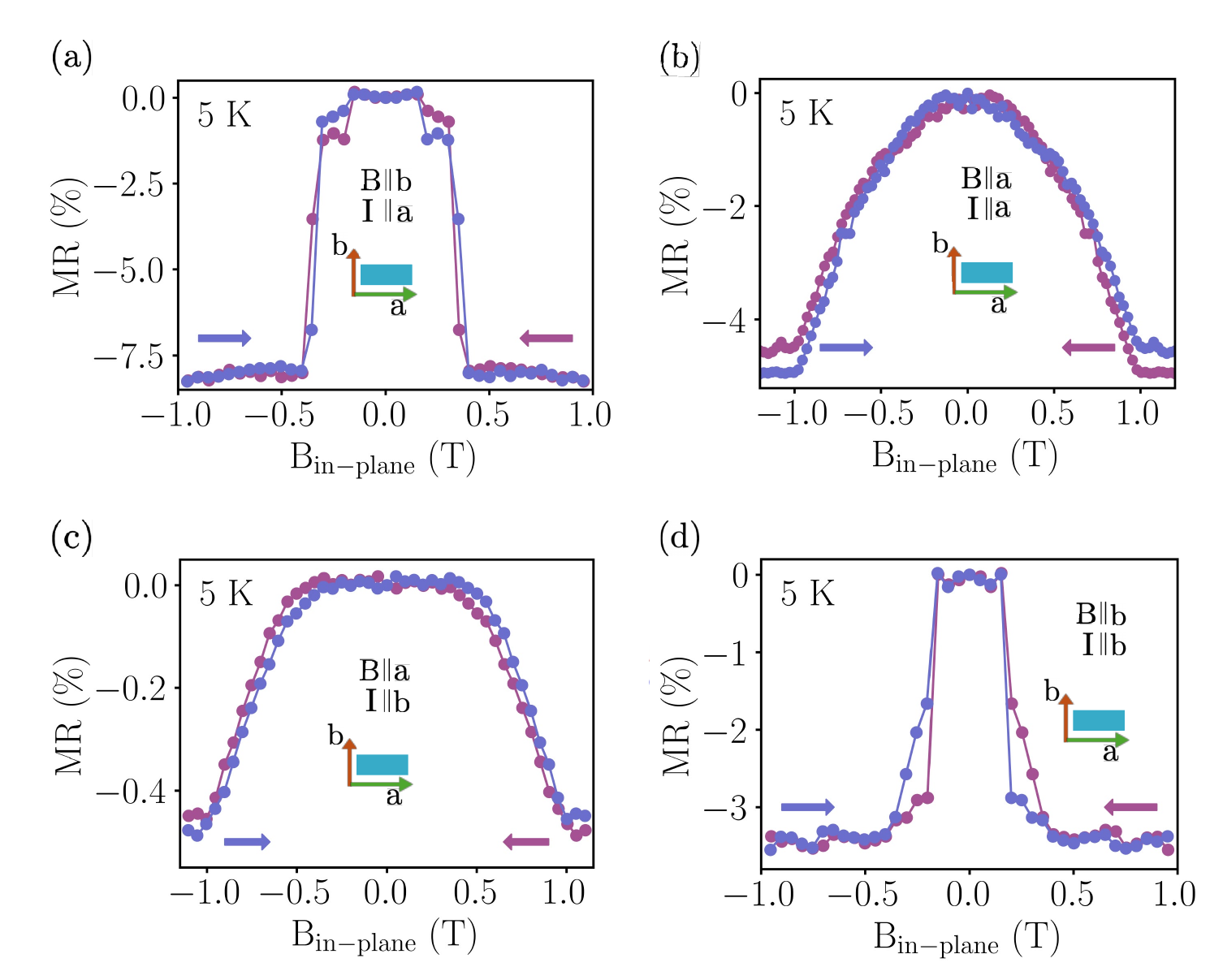}
\caption{Anisotropic magnetoresistance for different crystal axes with the magnetic field applied in-plane to the sample (thickness $\approx$ 80 nm). Current applied along $a-$axis; (a) B $ \perp$ I, (b)B $ \parallel$ I. Current along $b-$axis; (c) B $ \perp$ I, (d) B $ \parallel$ I }
\label{Fig3}
\end{figure}
We further performed in-plane MR measurements at 5~K with the current and magnetic field applied along the crystallographic $a$- and $b$-axes. As shown in Fig.~\ref{Fig3}, the MR is negative for all measurement configurations. Below the N\'eel temperature ($T_N$) and in zero magnetic field, the system is in an antiferromagnetic (AFM) state with antiparallel spin alignment, resulting in a higher electrical resistance. Upon application of a magnetic field, the system undergoes a field-induced transition from the AFM to the ferromagnetic (FM) state, characterized by parallel spin alignment and a corresponding reduction in resistance. Consequently, a negative MR is observed in all configurations. A clear hysteresis associated with the AFM--FM transition is present in the MR curves for all field and current orientations.

Figures~\ref{Fig3}(a) and \ref{Fig3}(b) show the MR measured with current applied along the $a$-axis and the magnetic field oriented perpendicular and parallel to it, respectively. In Fig.~\ref{Fig3}(a), the MR exhibits a sharp decrease near 0.5~T and subsequently reaches a saturation value of approximately $-8\%$. This abrupt change in MR reflects the two-step transition in magnetization when the field is applied along the $b$-axis, consistent with the $M$--$B$ behavior shown in Fig.~\ref{Fig1}(d). In contrast, in Fig.~\ref{Fig3}(b), the MR decreases more gradually and saturates at approximately $-5\%$ around 1~T, in agreement with the magnetization response for a field applied along the $a$-axis. In both cases, the magnetic field ($B_{sat}$) at which the MR saturates coincides with the field required to drive the system from an unpolarized AFM state to a fully spin-polarized FM state.

When the current is applied along the $b$-axis, the MR reaches approximately $-0.5\%$ and $-3.5\%$ for magnetic fields applied perpendicular and parallel to the current, respectively, as shown in Figs.~\ref{Fig3}(c) and \ref{Fig3}(d). The saturation field $B_{\mathrm{sat}}$ is identical to that observed previously for current applied along the $a$-axis for both field orientations. Consequently, from all measurement configurations, we infer that $B_{\mathrm{sat}}$ is governed by the direction of the applied magnetic field and is independent of the current direction. In contrast, the current direction determines the magnitude of the MR with respect to the applied magnetic field.  Overall, in-plane measurements driven magnetoresistance substantiate the presence of axial magnetic anisotropy in CrSBr [refer to Table 1 below].

\begin{table}[h!]
\centering

\begin{tabular}{|c|c|c|c|}
\hline
\textbf{I vs axis} & \textbf{B vs axis} & \textbf{B$_{sat}$(T)} & \textbf{MR$(\%)$ } \\ \hline 
I $ \parallel a$ & B $\perp a $ or  $B \parallel b$ & 0.45& -8  \\ \hline
I $ \parallel a$ & B $\parallel a $ & 1.0 & -5  \\ \hline
I $ \parallel b$ & B $\perp b $ or  $B \parallel a$ & 1.0 & -0.5  \\ \hline
I $ \parallel b$ & B $\parallel b $ & 0.45 & -3.75 \\ \hline
\end{tabular}
\caption{In-plane MR for all the configurations obtained from plots in Fig.~\ref{Fig3}}
\label{tab:angle_dependent_MR}

\end{table}

Our observations can be summarized as follows. The magnitude of MR exhibits hysteresis for all measured configurations, indicating a finite net magnetization for any in-plane field orientation. The MR reaches its maximum magnitude ($\approx$ 8\%) when the magnetic field and the current are mutually perpendicular, with the field applied along the crystallographic b-axis [fig. \ref{Fig3} (a)]. The hysteresis, however, is weak: the separation between the forward and reverse field sweeps is limited to approximately 0.3–0.4~T. This field scale is consistent with the magnetization measurements, which show magnetization reversal occurring in the same range. Furthermore, when current and field are parallel to each other in this configuration of magnetic field w.r.t. crystallographic axes[fig. \ref{Fig3} (d)], the MR drops to $\approx$~-3.5 \%. Opposite trend is observed when the experiment is performed with magnetic field set parallel to a-axis; MR is $\approx$~-5\% when current and field are parallel, MR is $\approx$~-0.5\% when current and field are perpendicular to each other. These findings suggest that the spin configuration prefers being oriented along the b-axis.  Since MR is a dimension-less quantity, geometry of the electrodes can be neglected for this comparison. Furthermore, the configuration where current being parallel to a-axis yields larger magnitude of MR compared to the configuration where current is perpendicular to a-axis.  This preferential MR w.r.t. crystallographic direction is a hallmark of anisotropic Fermi surface, even in the sample plane. We tabulate the magnitude of MR values (table-I) for a comprehensive reference. 

In this work, we investigated the magnetoresistance of exfoliated CrSBr flakes, exploiting the intrinsically anisotropic Fermi surface of this layered magnetic semiconductor. By performing systematic transport measurements along distinct crystallographic axes and under controlled orientations of the applied magnetic field, we demonstrate a pronounced anisotropy in the magnetoresistance response. These results establish magnetoresistance—despite its dimensionless and experimentally simple nature—as a highly sensitive electronic probe of Fermi surface anisotropy. More broadly, our findings suggest that angular- and axis-resolved magnetoresistance measurements provide a versatile and non-invasive methodology to quantify electronic anisotropy in low-dimensional and correlated materials. This approach can be readily extended to other anisotropic quantum materials, offering a practical route to characterize direction-dependent transport and spin–charge coupling, with direct implications for the rational design and optimization of anisotropic spintronic and magnetoelectronic devices.

\noindent Supplementary Material\\
The Supplementary Materials provide structural characterization data for an additional sample and along additional measurements.
\vspace{0.5cm}\\
\noindent DS thanks IISc start-up grant, Ministry of Electronics and Technology (MeiTy), Govt. of India, Indian Space Research Organization (ISRO), Govt. of India for funding. Authors are grateful for funding from INOX Airproducts and INOXCVA for funding via CSR grants. DS thanks support from Infosys Foundation, Bangalore. PB thanks Anusandhan National Research Foundation, National Postdoctoral fellowship (PDF/2023/000444) for financial support. Authors are grateful to micro and nano characterization facility, CeNSE, IISc and national nano-fabrication facility, CeNSE, IISc both supported by Government of India,  for facilities usage. AS thanks Science and Engineering Research Board (now Anusandhan National Research Foundation) - Core Research grant (CRG/2022/004311) and University of Hyderabad for financial support. \\
\vspace{0.5cm}\\
\noindent Data Availability\\
The data that supports the findings of this study are available from the corresponding author upon reasonable request.

\bibliography{04_references}
\end{document}


\title{Supplementary Information for the article \\ Crystallographic Orientation-Dependent Magnetotransport in the Layered Antiferromagnet CrSBr}

\author{Naresh Shyaga}
\affiliation{Centre for Nanoscience and Engineering, Indian Institute of Science, Bengaluru 560012, Karnataka, India} 
\author{Pankaj Bhardwaj}
\affiliation{Centre for Nanoscience and Engineering, Indian Institute of Science, Bengaluru 560012, Karnataka, India} 
\author{Rajib Sarkar}
\affiliation{Centre for Nanoscience and Engineering, Indian Institute of Science, Bengaluru 560012, Karnataka, India} 
\author{Jagadish Rajendran}
\affiliation{Centre for Nanoscience and Engineering, Indian Institute of Science, Bengaluru 560012, Karnataka, India}

\author{Abhiram Soori}
\affiliation{School of Physics, University of Hyderabad, Prof. C. R. Rao Road, Gachibowli, Hyderabad -- 500046, India} 
\author{Dhavala Suri}
\email{dsuri@iisc.ac.in}
\affiliation{Centre for Nanoscience and Engineering, Indian Institute of Science, Bengaluru 560012, Karnataka, India} 

\maketitle
\section{\NoCaseChange{Temperature \& Magnetic Field-dependent Magnetic properties of CrSBr crystal}}
\noindent We performed susceptibility vs temperature measurements at various fields, as shown in Fig. S1 (a) and (c) for both crystal axes $a-$ and $b-$axes, respectively. Both axes exhibit a change from the paramagnetic (PM) state to antiferromagnetic(AFM) phase below approxmiately 1~T and 0.5~T, respectively. Beyond these fields, curves showed the transition from PM to Ferromagnetic(FM) order.

\begin{figure}[!ht]
\centering
\includegraphics[width=1\textwidth]{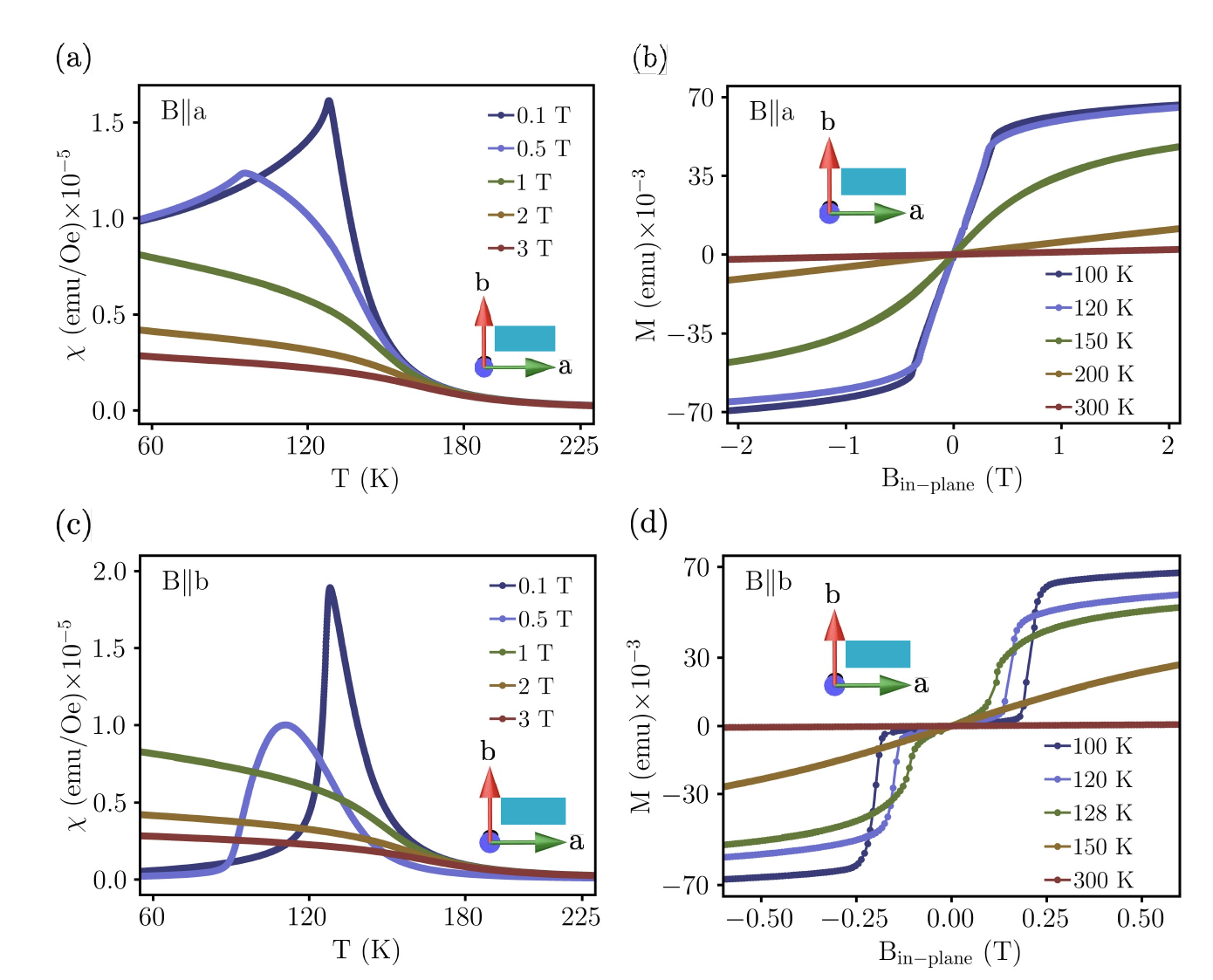} 
\label{Fig.MT-MH}
\noindent \justifying Fig.~S1:~(a) Magnetic field dependent susceptibility ($\chi$) vs temperature along the $a-$axis, (b) Temperature dependent Magnetization (M) vs magnetic field along the $a-$axis. (c)Susceptibility ($\chi$) vs temperature along the $b-$axis at different magnetic fields (d). Temperature-dependent M vs magnetic field along the $a-$axis.
\end{figure}
\noindent  The temperature dependent magnetization as function of applied field for both the a- and b-axes [ Fig.S1 (b) and (d)] shows a clear transformation from AFM state to FM state below the N\'eel Temperature($\mathrm{T_N}$). Above $\mathrm{T_N}$, the sample displays a paramagnetic state, reflected by the linear behaviour in the M vs B curves.

\newpage
\section{\NoCaseChange{Optical Images of \text{CrSBr}}}
\noindent CrSBr flakes were exfoliated onto SiO$_2$/Si substrate using the scotch tape method. We captured optical images at different locations on the sample, as depicted in the [Fig.S2 (a,b,c,d)]. We can clearly see that the exfoliated flakes are rectangular in shape owing to the crystal structure of CrSBr. Exfoliation produces flakes of different sizes, and they exhibit different colour contrasts due to their varying thicknesses. We will select the required flake and transfer it to the circular electrodes using dry transfer method.
\begin{figure}[!ht]
\centering
\includegraphics[width=1\textwidth]{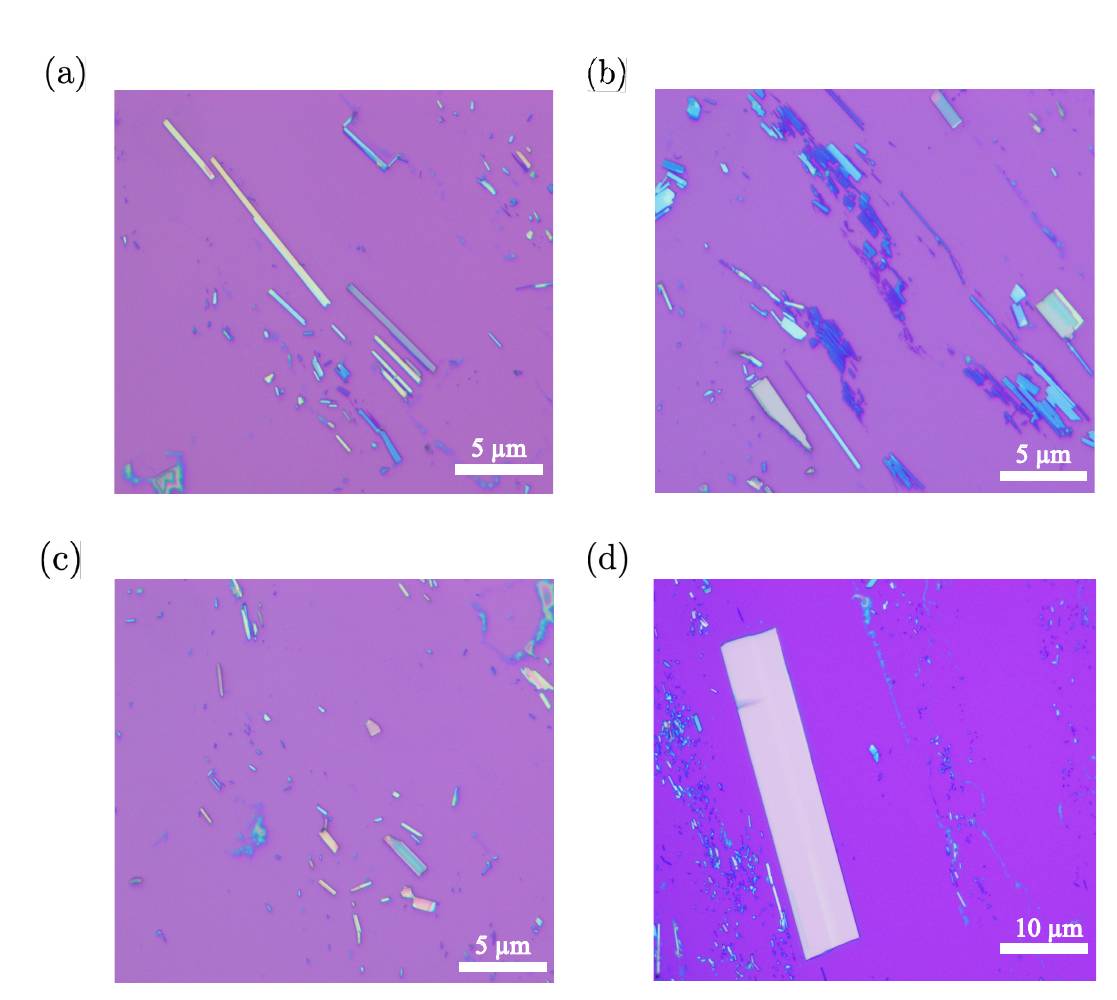} 
\label{Fig.optical}
\noindent \justifying Fig.~S2:~Optical micrographs (a),(b),(c), and (d) of exfoliated CrSBr flakes on SiO$_{2}$/Si substrate.
\end{figure}

\newpage
\section{\NoCaseChange{Temperature-dependent Resistance of CrSBr}}
\noindent Resistance as a function of temperature and its derivative  (dR/dT) is shown in Fig.S3 (a). The dR/dT has a local minimum [Fig.S3 (b)], indicating the transition from a paramagnetic to an antiferromagnetic state. This temperatuere can be ascribed to the N\'eel temperature ($\mathrm{T_N}$).
\begin{figure}[!ht]
\centering
\includegraphics[width=1\textwidth]{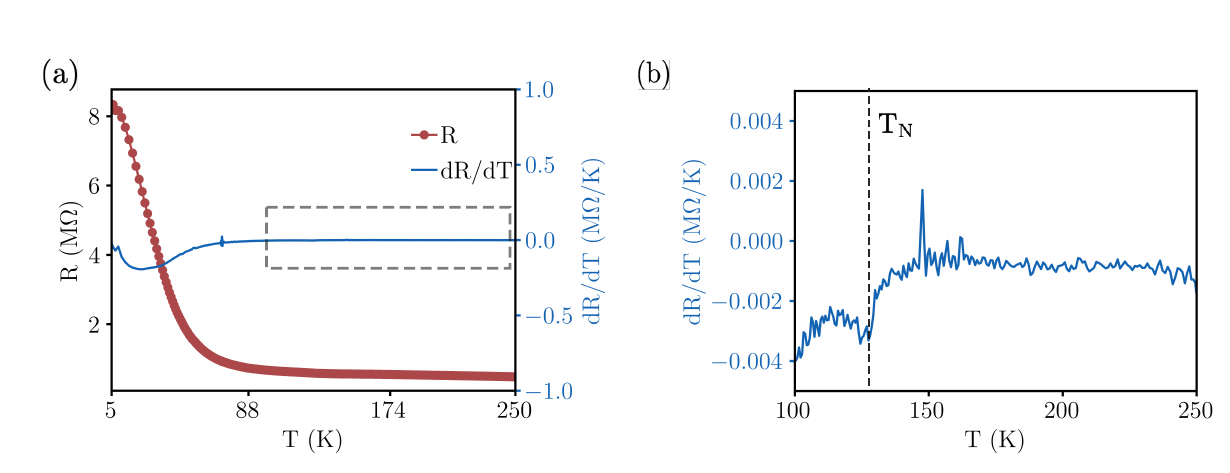} 
\label{Fig.RvsT}
\noindent \justifying Fig.~S3:~(a) Resistance (R) (left y-axis) as a function of temperature (T) and the derivative of resistance (dR) (right y-axis) w.r.t temperature. (b) dR/dT vs T from 100 K to 250 K. 
\end{figure}

\newpage
\section{\NoCaseChange{Thickness Measurements of CrSBr flakes}}
\noindent  We measured the thicknesses of the CrSBr flakes using atomic force microscopy. Fig.S4 (a) shows an optical image of the flake on the circular electrodes. We scanned at the edge of the flake, depicted in Fig. S4.(b) and found a thickness of about $\approx$ 210 $\pm$ 2 nm [Fig. S4.(c)]. Similarly,the thickness of the second flake was found to be  78 $\pm$ 2 nm, as shown in [Fig.S4(d-f)].
\begin{figure}[!ht]
\centering
\includegraphics[width=1\textwidth]{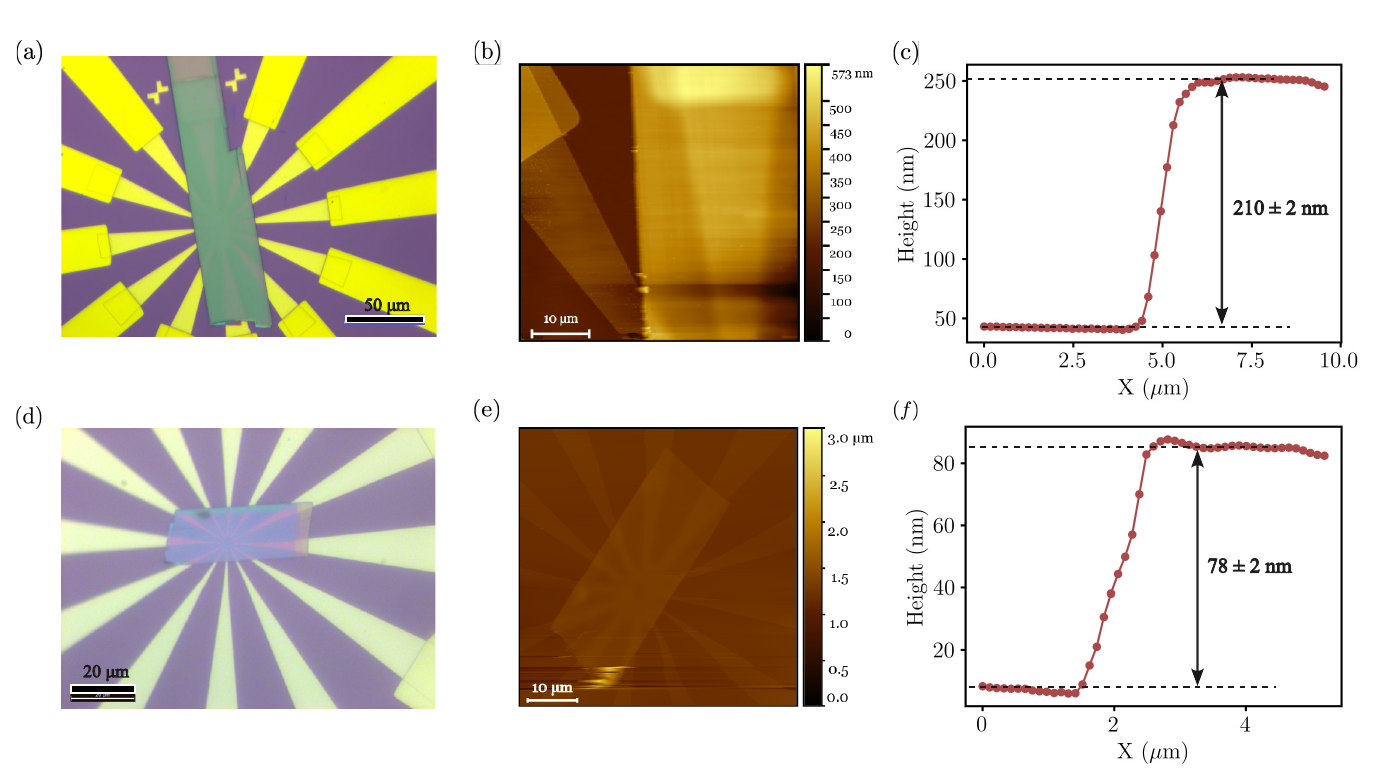} 
\label{Fig.AFM}
\noindent \justifying Fig.~S4:~Thicknesses of two flakes. (a)\&(d) Optical micrographs of CrSBr flakes on circular electrodes. (b)\&(e) Atomic force microscopic images of both the flakes. (c)\&(f) Height profile of both flakes.
\end{figure}
\newpage

